\documentstyle[epsfig,a4]{article}

\begin{document}

\begin{titlepage}

\centerline{\Large \bf Persistence Probabilities of}
\centerline{\Large \bf the German DAX and Shanghai Index}

\vskip 2.0truecm
\centerline{\bf F. Ren$^{1,2}$, B. Zheng$^{1,2}$,
H. Lin$^1$, L. Y. Wen$^1$ and S. Trimper$^2$}
\vskip 0.2truecm
\centerline{$^1$ Zhejiang University, Zhejiang Institute of Modern
Physics, Hangzhou 310027, P.R. China}
\centerline{$^2$ FB Physik,
Universit\"at -- Halle, 06099 Halle, Germany}
\vskip 2.5truecm

\abstract{We present a relatively detailed analysis of the
persistence probability distributions in financial dynamics.
Compared with the auto-correlation function, the persistence
probability distributions describe dynamic correlations non-local
in time. Universal and non-universal behaviors of the German DAX
and Shanghai Index are analyzed, and numerical simulations of some
microscopic models are also performed. Around the fixed point
$z_0=0$, the interacting herding model produces the scaling
behavior of the real markets.}

\vspace{0.5cm}

{\footnotesize PACS: 05.40.-a, 01.75.+m, 89.90.+n}

\vspace{0.3cm}

{\footnotesize Keywords: nonequilibrium kinetics, financial dynamics, critical dynamics}

\end{titlepage}

\baselineskip=24.0pt

\section{Introduction}

Recently, much attention of physicists has been drawn to dynamics
of financial markets. From the view of many-body systems,
interactions among agents and producers may generate long-range
temporal correlations in financial dynamics, and therefore result
in the so-called dynamic scaling behavior. To clearly observe the
scaling behavior, one needs to carefully investigate the dynamics
at the 'microscopic' level.

Fortunately, in the past years, it has been piled up large amount
of data in financial markets, especially those records of economic
indices in {\it minutes or seconds}. This allows a relatively
accurate analysis of the financial dynamics at the 'microscopic'
level. In 1995, Mantegna and Stanley had carefully analyzed the
data of a stock market index in US --- the Standard and Poor 500
\cite{man95}. The probability distribution $P(\Delta y,\Delta t)$
of the price change $\Delta y$ in a time $\Delta t$ obeys a
dynamic scaling form.

Inspired by the work of Mantegna and Stanley \cite{man95}, many
activities have been devoted to the study of financial markets
\cite{gha96,man96,sor96,fei96,sta98,lal99,ple99,lux99,egu00,sta01,gia01}.
More systematic investigation of the tails of $P(\Delta y,\Delta
t)$ and the volatility of price fluctuations in the Standard and
Poor 500 has been presented \cite{gop99,liu99}. In spite of {\it
absence} of the two-point correlation of the variations, the
volatility is long-range correlated. This is believed to be the
physical origin of the universal scaling behavior in financial
dynamics. On the other hand, different models and theoretical
approaches have been also developed to describe financial markets
\cite{lux99,gia01,cha97,sta99,sta00,con00,egu00,muz00,cha01,lou01,kra02,hua02}.
Among them, important examples are Minority games and percolation
models and their variants.

Up to date, stationary properties of financial indices and the
dynamic behavior local in time are mainly concerned. For full
understanding of financial markets, however, the dynamic behavior
{\it non-local} in time should be also very important in theory
and practice. An interesting example is the so-called persistence
probability, which has been systematically investigated in
non-equilibrium dynamics such as phase ordering dynamics and
critical dynamics. As time evolves, the persistence probability
decays by a power law characterized by a persistence exponent. It
is shown that the persistence exponent is in general {\it
independent of} other known critical exponents \cite{maj96a}. The
persistence exponent plays an important role in a variety of
systems and has been directly measured in experiments
\cite{maj99,der94,maj96,oer98,zhe98}.

The persistence probability is not only important in the
fundamental theory of the non-equilibrium dynamics, but also
helpful in thorough understanding of financial markets. The
persistence probability directly defined with the price of the
index $y(t')$, is first introduced in financial dynamics
\cite{zhe02}. The persistence exponent $\theta_p$ is estimated to
be $0.49(2)$. It is very close to that of a random walk, even
though the probability distribution $P(\Delta y,\Delta t)$ of the
financial index is significantly different from Gaussian. This
result should be related to the fact that $\Delta y(t') =
y(t'+\Delta t) - y(t')$ is short-range correlated in time.

Consequently, the persistence probability defined with the {\it
magnitude} of the variation $\Delta y(t')$ of the index is
investigated \cite{ren03}. Since the magnitude of the variation
$\Delta y(t')$ is long-range correlated in time, one expects a
non-trivial behavior of the persistence probability distribution.
Some preliminary results of this kind have been reported for the
German DAX in Ref. \cite{ren03}. In this paper, a more detailed
analysis on this topic will be presented, and further extended to
the case away from the fixed point. In addition to the German DAX,
the Shanghai Index in China is also carefully analyzed, and
universal and non-universal behaviors of two economic indices are
revealed. Although up to date not much work has been devoted to
the Chinese stock market, it should be rather interesting to study
the dynamic behavior of the Chinese market, in comparison with
that of the financial market in western countries. This may help
understand what is real universal in stock markets. Finally,
numerical simulations of an interacting herding model
\cite{zhe04b} and a Minority Game model \cite{cha01} are briefly
reported.

In Sec. 2, the persistence probability distribution is introduced
and investigated. In Sec. 3, it is generalized to the case away
from the fixed point. Finally, conclusions come in Sec. 4.

\section{Persistence probabilities}

\subsection{The German DAX and Shanghai Index}

Let us denote the value of a index at a certain time $t'$ as
$y(t')$, and the magnitude of the logarithm price change in a
fixed time interval $\Delta t$ as $Z(t',\Delta t) \equiv |\ln
y(t'+\Delta t)-\ln y(t')|$. We define the persistence probability
$P_+(t)$ ($P_-(t)$) as the probability that $Z(t'+\tilde t,\Delta
t)$ has always been above (below ) $Z(t',\Delta t)$ in time $t$,
i.e., $Z(t'+\tilde t,\Delta t) > Z(t',\Delta t)$ ($Z(t'+\tilde
t,\Delta t) < Z(t',\Delta t)$) for all $\tilde t < t$. The average
is taken over the time variable $t'$.

Here $Z(t',\Delta t)$ is defined as the magnitude of the variation
of $\ln y(t')$, slightly different from that in Ref. \cite{ren03}.
Taking $\ln y(t')$ in the definition of $Z(t',\Delta t)$, one may
eliminate possible effects of the background evolution of the
index. In Ref. \cite{ren03}, $Z(t',\Delta t)$ is defined as the
magnitude of the variation of $y(t')$, and therefore, the
background evolution should be directly removed in $y(t')$. Both
definitions of $Z(t',\Delta t)$ yield similar results, but that
with $\ln y(t')$ is slightly more fluctuating. In the present
paper, we adopt the definition with $\ln y(t')$ since the total
time interval of the Shanghai Index is too short for figuring out
the the background evolution.

The persistence probabilities describe the temporal correlation of
$Z(t',\Delta t)$ {\it non-local} in time, while the
auto-correlation function describes the temporal correlation local
in time. In general, the persistence probabilities provide
additional information to the auto-correlation function, and are
less fluctuating in the measurements. Therefore, they are
interesting and important for deep understanding of the financial
dynamics. For a random walk process, both $P_+(t)$ and $P_-(t)$
decay by a power law with a persistence exponent $\theta_p=1$.
Since $Z(t',\Delta t)$ in financial dynamics is long-range
correlated in time, one expects a non-trivial dynamic behavior.

\begin{figure}[p]\centering
\epsfysize=7.cm
\epsfclipoff
\fboxsep=0pt
\setlength{\unitlength}{0.7cm}
\begin{picture}(9,9)(0,0)
\put(-2,0.0){{\epsffile{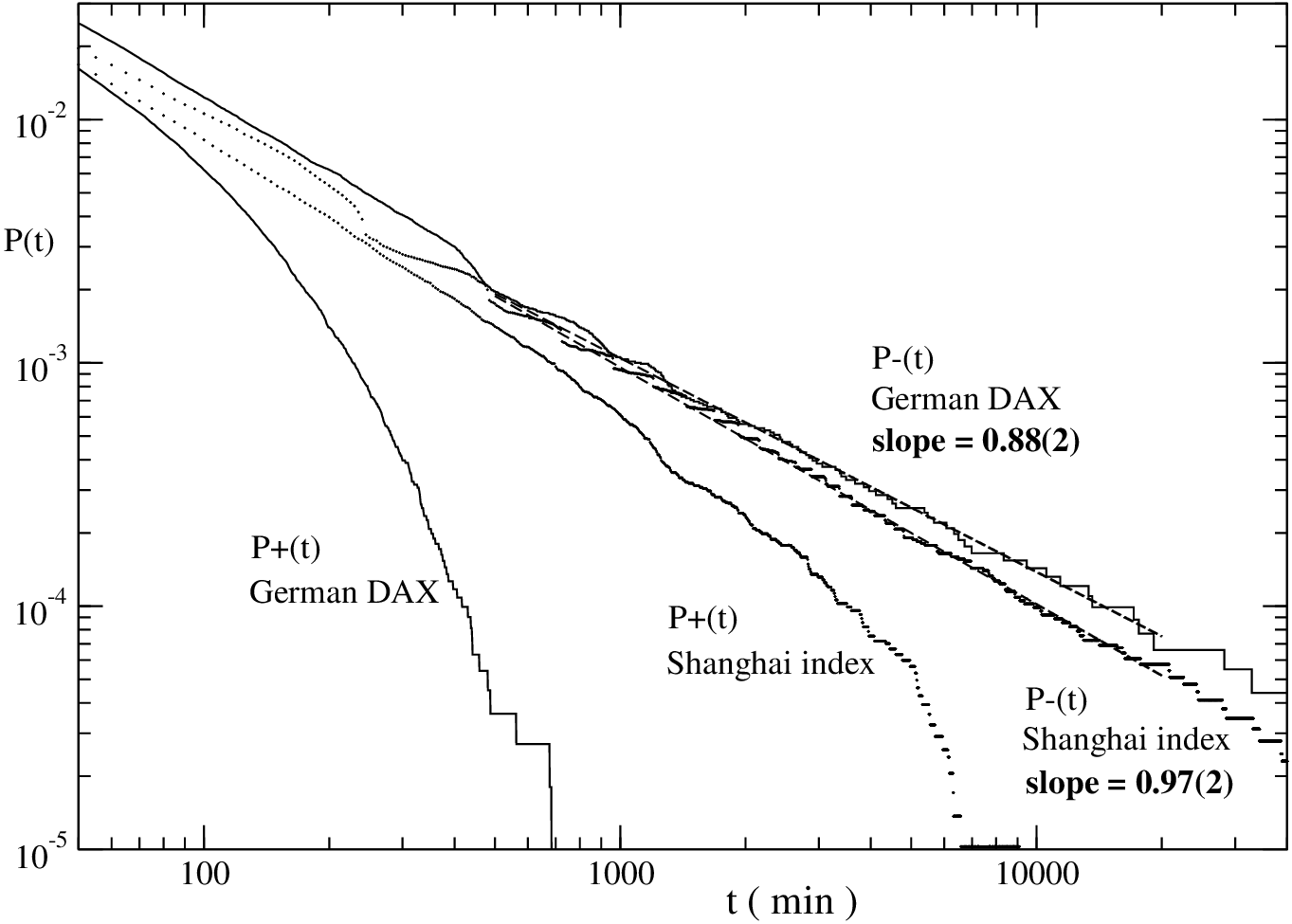}}}
\end{picture}
\caption{Persistence probabilities displayed in log-log scale.
Curves of the German DAX (solid lines) are obtained with the
minute-to-minute data and $\Delta t = 1 \ minute$, while those of
the Shanghai Index (cross lines) are calculated with the data
recorded every 5 minutes and $\Delta t = 5 \ minutes$. Dashed
lines fitted to the curves show the power-law fits.} \label{f1}
\end{figure}

We first perform the measurements using the minute-to-minute data
of the German Dax from December $1993$ to July $1997$. The total
number of the records during this period of time is about $350\
000$. In the measurements, we take $\Delta t=1$ minute. In
Fig.~\ref {f1}, we plot the persistence probabilities $P_+(t)$ and
$P_-(t)$ on a log-log scale. It is obvious that $P_-(t)$ obeys a
power law up to four orders of magnitude, while $P_+(t)$ decays to
zero rather fast. Compared with the auto-correlation function
calculated with the same data, $P_+(t)$ and $P_-(t)$ are much less
fluctuating. Different behaviors of $P_+(t)$ and $P_-(t)$ indicate
the high-low asymmetry in the time series of $y(t')$. For
$P_-(t)$, we assume a power law
\begin{equation}
P_-(t) \sim t^{-\theta_p}, \label{e10}
\end{equation}
and $\theta_p$ is the so-called persistence exponent.

Carefully looking at the curve of $P_-(t)$ of the German DAX, we
observe a quasi-periodic dropping in the first $2000$ minutes. The
period is roughly one working day, i.e., about $350$ to $480$
minutes in those years. This behavior should be traced back to the
disconnection of the index between two successive days.
$Z(t',\Delta t=1 \ min)$ at the last minute of a day is much
bigger than those during the day. This disconnection affects the
behavior of $P_-(t)$ in the first periods of time. When one
measures the slope in a time interval $[500,20000]$, the
persistence exponent of $P_-(t)$ is $\theta_p=0.88(2)$, clearly
different from $\theta_p=1$ for a random walk. This indicates that
$Z(t',\Delta t)$ is indeed long-range correlated in time.

For comparison, we have also performed the measurements with the
data of the Shanghai Index from January $1998$ to July $2003$. The
time interval between successive records is $5$ minutes. As shown
in Fig.~\ref {f1}, similar to the case of the German DAX, $P_-(t)$
obeys a power law and $P_+(t)$ decays faster. A quasi-periodic
dropping of $P_-(t)$ in early times is also observed, but the
period now is less than $300$ minutes, because the working day in
China is about or less than five hours in those years. If one
measures the the slope of the curve in the time interval
$[500,20000]$, the persistence exponent $\theta_p=0.97(2)$ is
obtained.

In general, the behavior of $P_+(t)$ is not expected to be
universal, since it is not power-law-like, and it depends
essentially on $\Delta t$ in calculating $Z(t',\Delta t)$.  In
Fig.~\ref {f1}, $P_+(t)$ of the Shanghai Index decays slower than
that of the German DAX, just because $\Delta t = 5 \ mins$ for the
Shanghai Index and $\Delta t = 1 \ min$ for the German DAX. As
$\Delta t$ increases, the high-low asymmetry is continuously
weakened. This tendency can be clearly seen in the analysis of the
daily data in the next subsection.

Since $P_-(t)$ obeys a power-law, one may expect its behavior is
universal. However, it is somewhat puzzling that the persistence
exponent $\theta_p=0.97(2)$ of the Shanghai Index is so close to
$\theta_p=1.0$, the value of a random walk. This point will
further be investigated in the next subsection.

\subsection{The daily data and minute-to-minute data}

\begin{figure}[p]\centering
\epsfysize=7.cm
\epsfclipoff
\fboxsep=0pt
\setlength{\unitlength}{0.7cm}
\begin{picture}(9,9)(0,0)
\put(-2,-0.5){{\epsffile{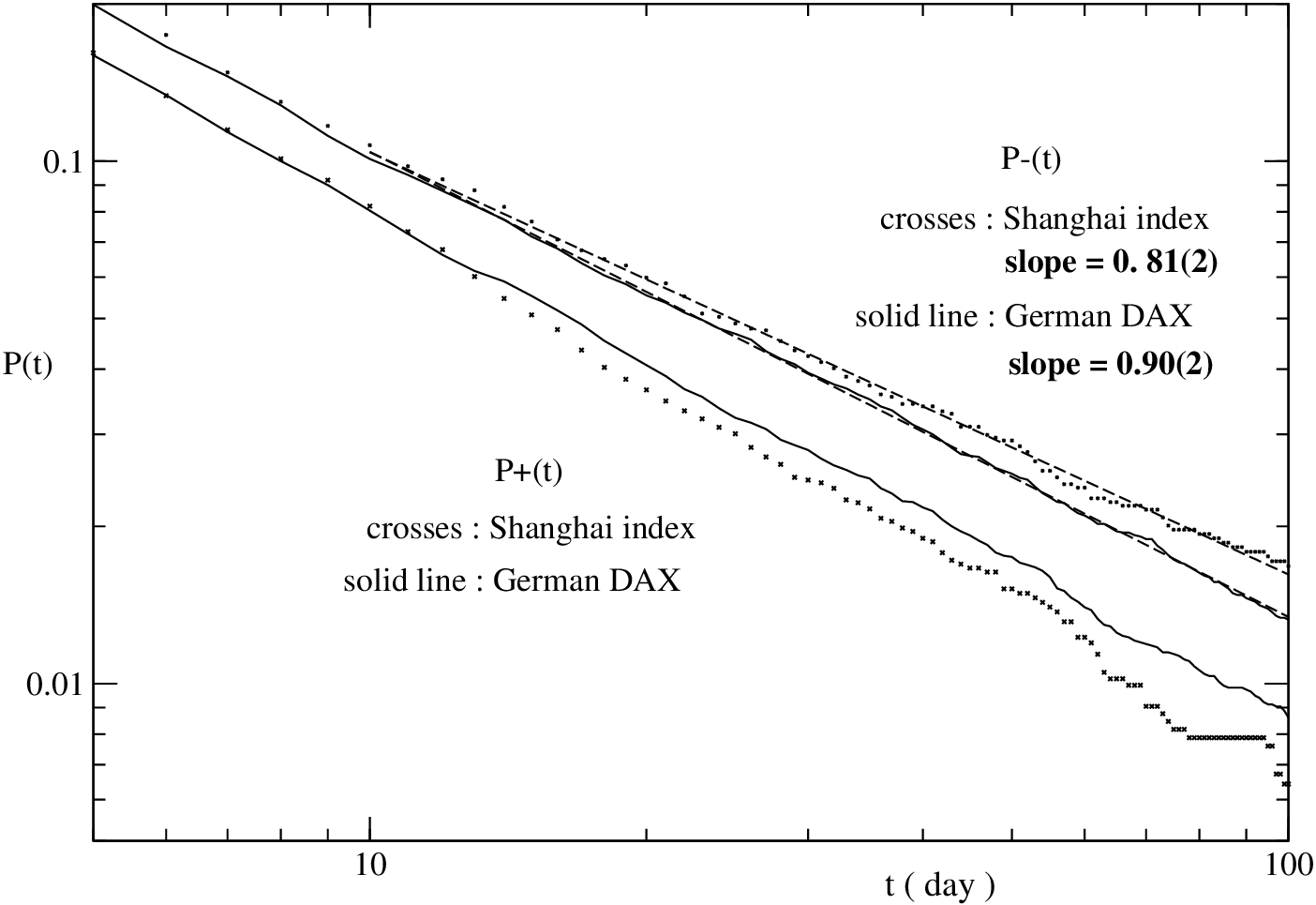}}}
\end{picture}
\caption{Persistence probabilities of the daily data in log-log
scale. Curves are obtained with $\Delta t = 1 \ day$.} \label{f2}
\end{figure}

To push forward our investigation starting with the
minute-to-minute data in the previous subsection, we now perform
the measurements with the daily data of the German DAX from
October $1959$ to January $1997$, and those of the Shanghai Index
from December $1990$ to December $2003$. In Fig.~\ref {f2},
$P_+(t)$ and $P_-(t)$ of the daily data with $\Delta t = 1 \ day$
are plotted on a log-log scale. In Fig.~\ref {f2}, we observe:

i) For both indies, compared with the case of the minute-to-minute
data, it seems $P_+(t)$ comes closer to a power law, but still
remains different from $P_-(t)$. This confirms that the behavior
of $P_+(t)$ depends on $\Delta t$.

ii) For both indies, at least up to 100 day, one observes a
power-law behavior of $P_-(t)$. The persistence exponent is
estimate to be $\theta_p=0.90(2)$ for the German DAX, and
$\theta_p=0.81(2)$ for the Shanghai Index.

\begin{figure}[p]\centering
\epsfysize=7.cm \epsfclipoff \fboxsep=0pt
\setlength{\unitlength}{0.7cm}
\begin{picture}(9,9)(0,0)
\put(-2,-0.5){{\epsffile{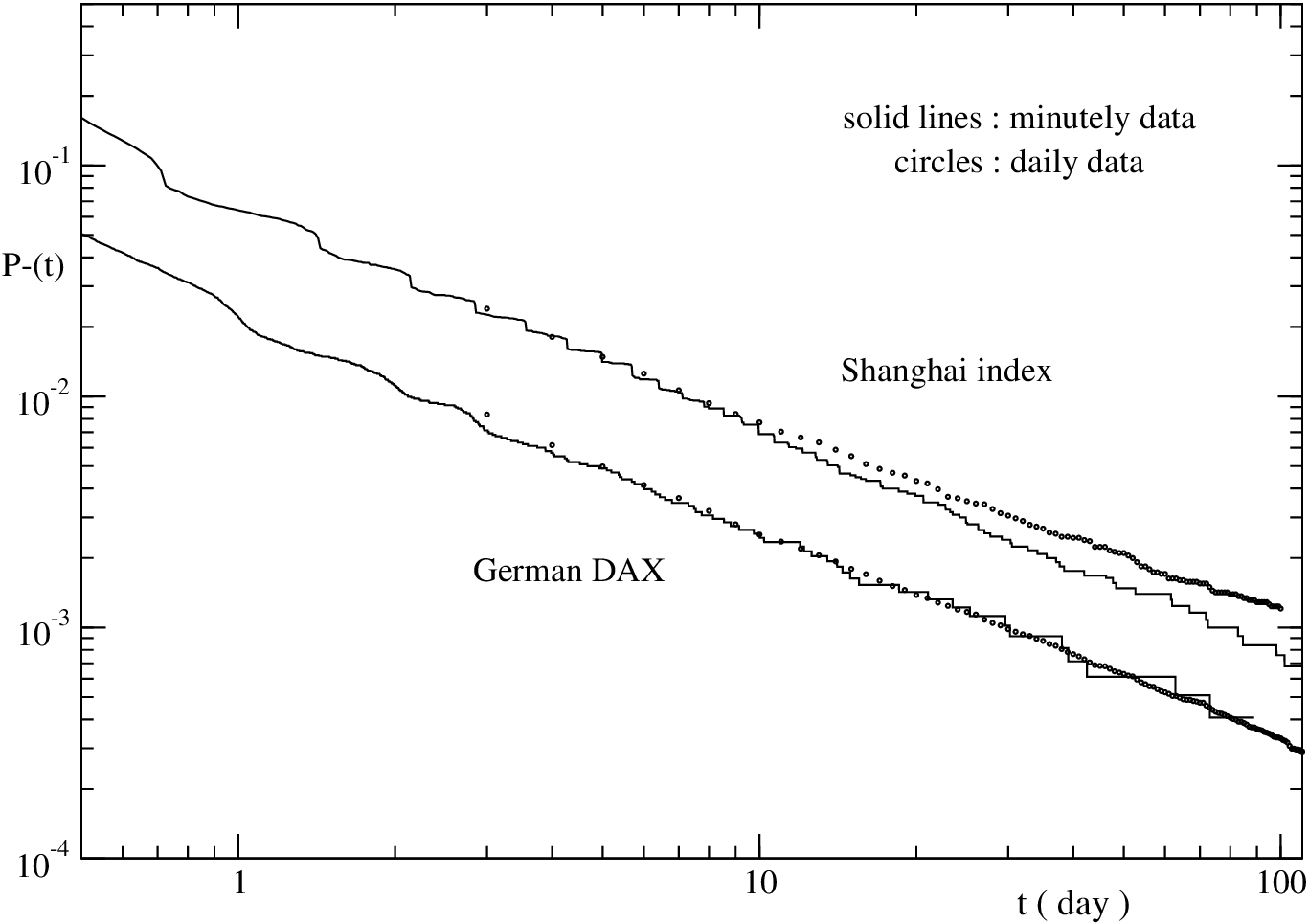}}}
\end{picture}
\caption{Comparison of the persistence probabilities measured with
the daily data and minutely data. The curves are taken from Figs.
\ref {f1} and \ref {f2}. The time unit for the curves of the
minutely data is $1 \ day = 400 \ mins$ for the German DAX and $1
\ day = 300 \ mins$ for the Shanghai Index.}
 \label{f3}
\end{figure}

The persistence exponent $\theta_p=0.90(2)$ measured with the
daily data of the German DAX is consistent within the errors with
$\theta_p=0.88(2)$ with the minute-to-minute data. This indicates
that the behavior of $P_-(t)$ is 'universal' for the German DAX
from $\Delta t =1 \ min$ to $1 \ day$. However, $\theta_p=0.81(2)$
measured with the daily data of the Shanghai Index is very
different from $\theta_p=0.97(2)$ measured with the
minute-to-minute data. To see this clearly, we rescale the time
unit in Fig. \ref {f1} to a working day, and compare the curves
with those in Fig. \ref {f2}. This is shown in Fig. \ref {f3}.

The difference between the minute-to-minute data and daily data of
the Shanghai Index should not come from the disconnection between
two working days, since this disconnection exists also for the
German DAX. We conjecture that the minute-to-minute data of the
Shanghai Index may be disturbed by many 'microscopic interactions'
from the environment, and therefore, the dynamic behavior is
different from that of the daily data. To confirm our conjecture,
we perform additional calculations for the minute-to-minute data
of the Shanghai Index with $\Delta t =300 \ min$, which is about a
working day. In other words, we uniformly select one datum from
the records of every $300$ minutes. Sometimes, we call this
procedure 'renormalization'.

\begin{figure}[p]\centering
\epsfysize=7.cm \epsfclipoff \fboxsep=0pt
\setlength{\unitlength}{0.7cm}
\begin{picture}(9,9)(0,0)
\put(-2,-0.5){{\epsffile{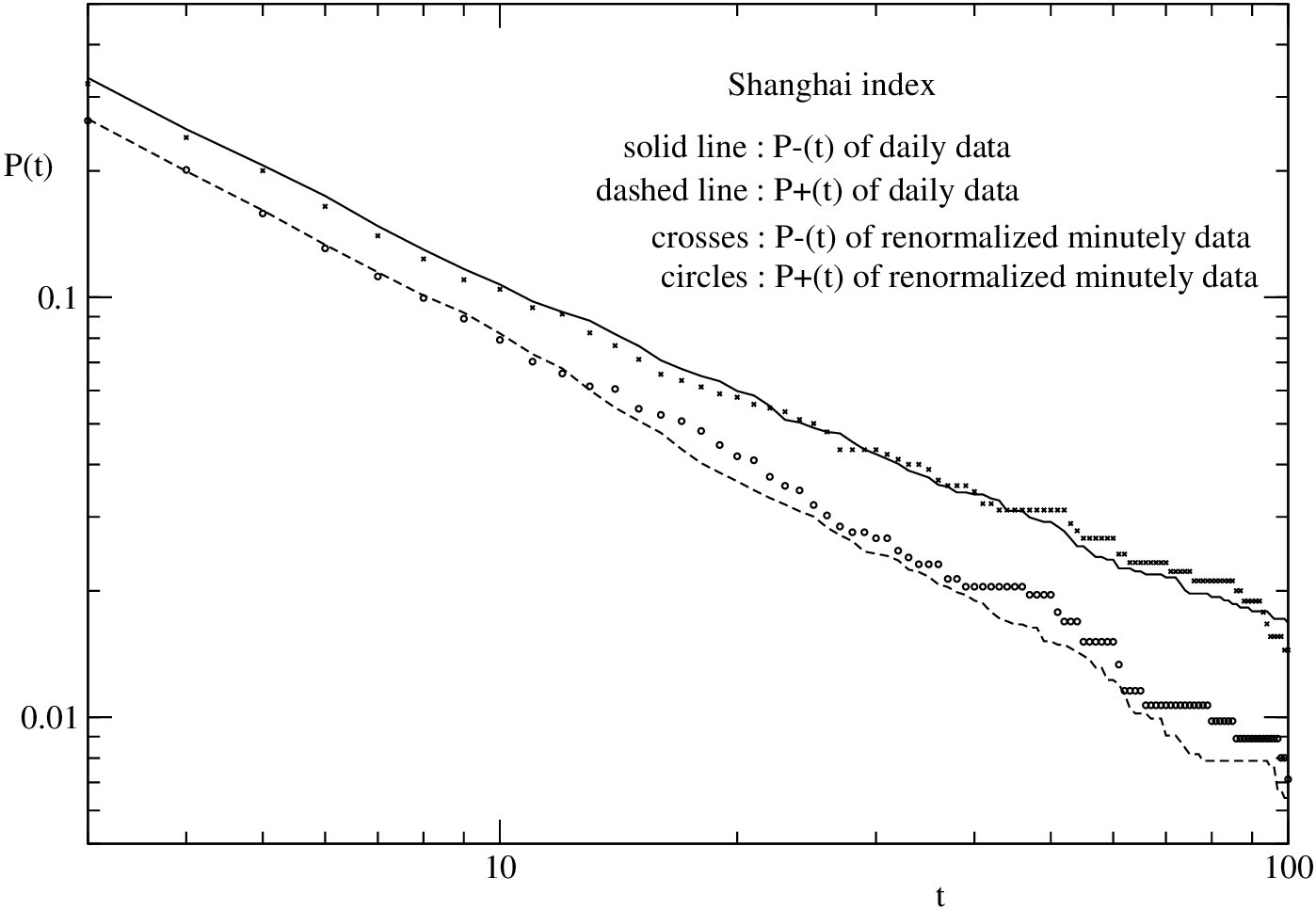}}}
\end{picture}
\caption{Persistence probabilities of the Shanghai Index in
log-log scale. The curves of the minutely data are calculated with
$\Delta t = 300 \ mins$, i.e., with the data selected every $300 \
mins$.} \label{f4}
\end{figure}

The persistence probabilities obtained with $Z(t',\Delta t = 300 \
mins)$ of the Shanghai Index are displayed in Fig. \ref {f4}, in
comparison with those of the daily data. Interestingly, both the
minute-to-minute data and daily data tend to exhibit similar
dynamic behaviors and give a same persistence exponent. It seems
that after we renormalize the time scale of the minute-to-minute
data to $\Delta t =300 \ mins$, the effects of the microscopic
interactions from the environment are eliminated. In this sense,
the behavior of $P_-(t)$ is also universal for the Shanghai Index.

Similar calculations with $\Delta t =400 \ mins$ have been also
performed for the minute-to-minute data of the German DAX, and the
results are in agreement with those of the daily data.

Finally, summarizing both measurements with the minute-to-minute
data and daily data, the persistence exponent is $\theta_p = 0.89
(2)$ for the German DAX and $\theta_p = 0.80 (2)$ for the Shanghai
Index. The different values of the persistence exponent for both
indices indicates that the financial markets in Germany and China
may belong to different universality classes.

\subsection{Modelling the real markets}

In this subsection, we present numerical simulations of some
microscopic models of the financial markets for comparison.

The herding model proposed by Eguiluz and Zimmermann \cite{egu00}
is simple and interesting, in which the clusters form themselves
dynamically during the time evolution, but it does not provides a
long-range temporal correlation. Therefore, a feedback interaction
is introduced: the rate of transmission of information at time
$t'$ depends on the price change at time $t'-1$. Then, volatility
clustering is generated \cite{zhe04b,zhe04}.

\begin{figure}[p]\centering
\epsfysize=7.cm \epsfclipoff \fboxsep=0pt
\setlength{\unitlength}{0.7cm}
\begin{picture}(9,9)(0,0)
\put(-2,-0.5){{\epsffile{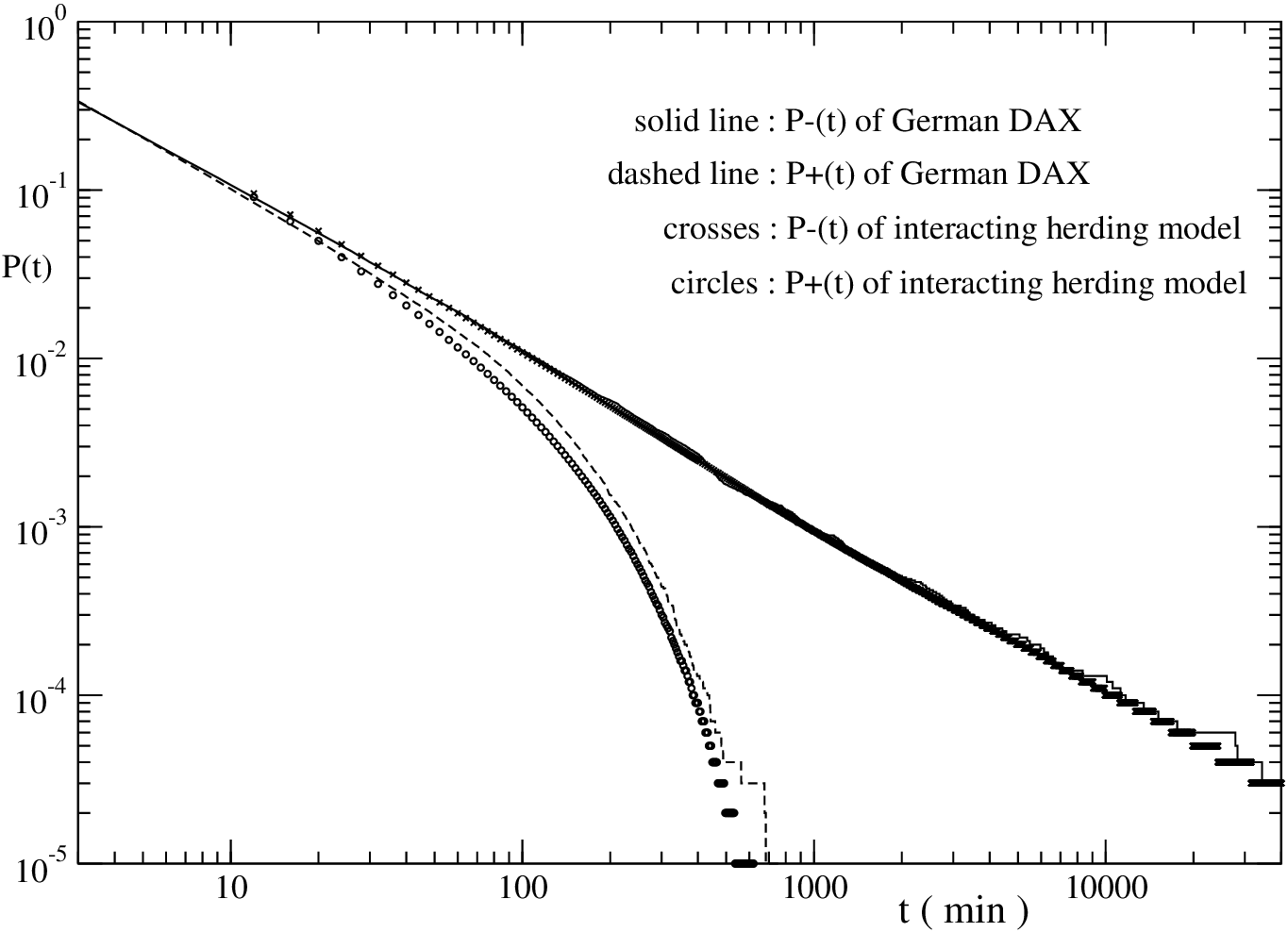}}}
\end{picture}
\caption{Persistence probabilities of the minutely data of the
German DAX are compared with those of the interacting herding
model.} \label{f5}
\end{figure}

In Fig.~\ref {f5}, $P_{+}(t)$ and $P_{-}(t)$ of the interacting
herding model are compared with those of the minute-to-minute data
of the German DAX . After taking an average of the time series
$y(t')$ generated by simulations over every $4$ time steps, the
interacting herding model reproduces nicely the minute-to-minute
behavior of the German DAX for both $P_{+}(t)$ and $P_{-}(t)$.

\begin{figure}[p]\centering
\epsfysize=7.cm \epsfclipoff \fboxsep=0pt
\setlength{\unitlength}{0.7cm}
\begin{picture}(9,9)(0,0)
\put(-2,-0.5){{\epsffile{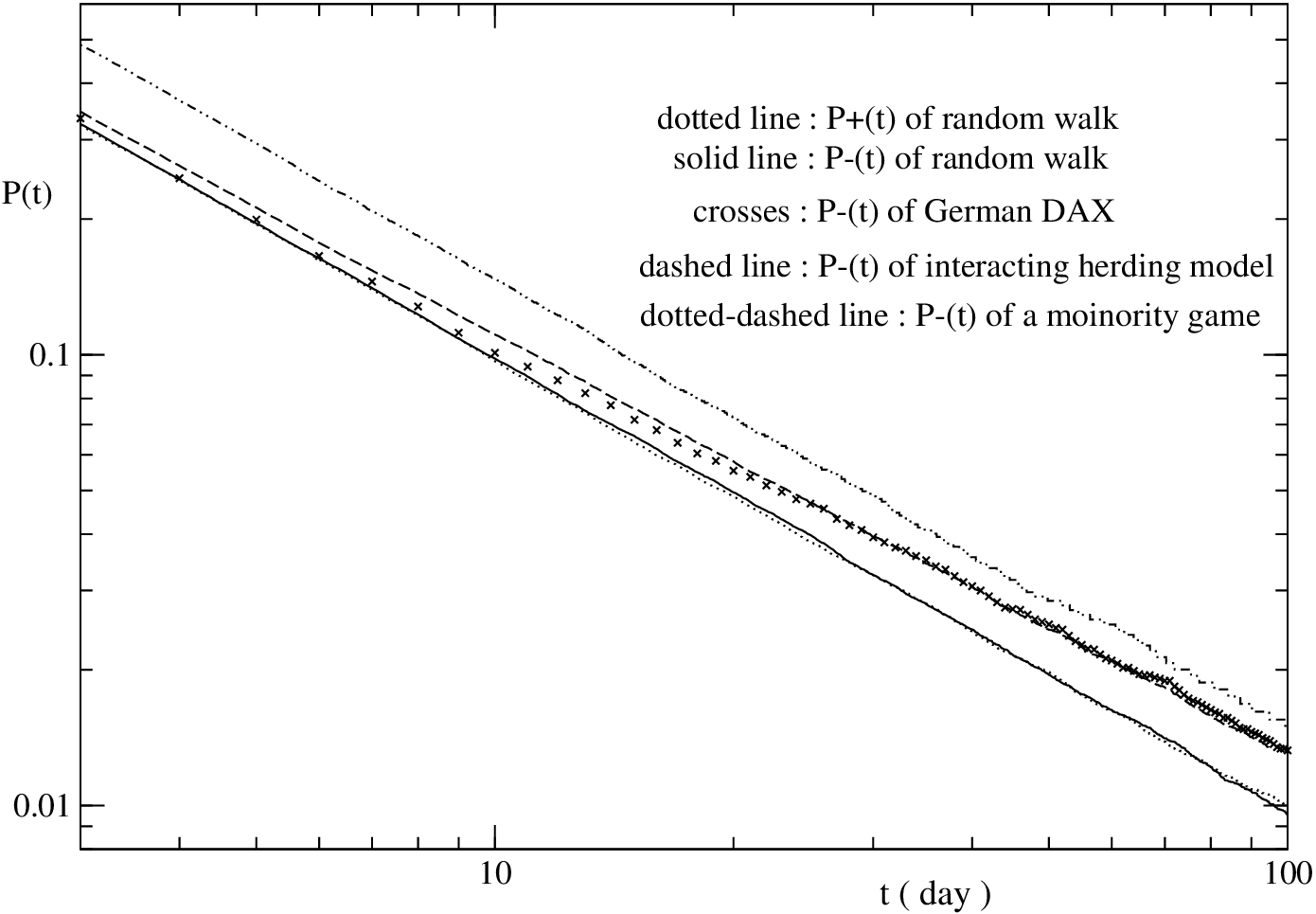}}}
\end{picture}
\caption{Persistence probabilities of the daily data of the German
DAX are compared with those of the random walk, the Minority Game
and the interacting herding model.} \label{f6}
\end{figure}

In Fig.~\ref {f6}, we compare the persistence probabilities of the
daily data of the German DAX, a random walk, a Minority Game with
an inactive strategy \cite{jef00,cha01} and the interacting
dynamic herding model. $P_{+}(t)$ and $P_{-}(t)$ of the random
walk, and $P_{+}(t)$ (not in the figure) and $P_{-}(t)$ of the
Minority Game all show a power-law behavior with a slope of $1.0$,
obviously different from those of the experimental data. The curve
of $P_{-}(t)$ of the interacting herding model is consistent well
with that of the experimental data.

We have also investigated the persistence probabilities of some
other models, it seems not so easy to reproduce the behavior of
the real markets. Even though the Minority Game with an inactive
strategy produces most stylized facts of the financial markets,
including the power-law behavior of the auto-correlation function,
it does not offer a correct behavior for the persistence
probabilities. The interacting herding model does produce the
high-low asymmetry of the minute-to-minute data of the German DAX,
and also a nontrivial persistence exponent $\theta_p=0.93(2)$. But
the 'renormalization' does not yield the results of the daily
data, i.e, it is not possible to produce the curves of $P_{+}(t)$
and $P_{-}(t)$ of the daily data of German DAX in Fig.~\ref {f2}
by selecting one datum from every some hundred time steps. (The
curve of $P_{-}(t)$ in Fig.~\ref {f6} is obtained only by
rescaling the time unit in Fig.~\ref {f5} to a working day, to be
compared with that of the daily data of the German DAX.) The
underlying reason may be simple: the interacting herding model
generates only the temporal correlation for $|\Delta
y(t')=y(t'+1)-y(t')|$, the sign of $\Delta y(t')=y(t'+1)-y(t')$ is
randomly given, and therefore, selecting one datum of $y(t')$ from
every some hundred time steps will lose most information of the
time correlation.

\section{General persistence probabilities}

For further understanding the dynamic behavior of financial
markets, we introduce more general persistence probability
distributions. Assuming that $z_0$ is a real positive number, we
define the generalized persistence probability $P_+(t,z_0)$
($P_-(t,z_0)$) as the probability that $Z(t'+\tilde t,\Delta t)$
has never been down (up) to $Z(t',\Delta t)-z_0$ ($Z(t',\Delta
t)+z_0$) in time $t$, i.e., $Z(t'+\tilde t,\Delta t) > Z(t',\Delta
t)-z_0$ ($Z(t'+\tilde t,\Delta t) < Z(t',\Delta t)+z_0$) for all
$\tilde t < t$. At $z_0=0$, $P_+(t,z_0)$ and $P_-(t,z_0)$ coincide
with the persistence probabilities defined in the previous
section.

For $P_{-}(t,z_0)$, we may write down a generalized dynamic
scaling form
\begin{equation}
P_{-}(t,z_0) = t^{-\theta_p}F_{-}(t^{\alpha_-}z_0), \label{e20}
\end{equation}
$\alpha_-$ is an exponent describing the scaling behavior of
$z_0$. Obviously the power-law behavior in Eq. (\ref {e10}) is
recovered at $z_0=0$. In this sense, $z_0=0$ is a fixed point in
the dynamic system

\begin{figure}[p]\centering
\epsfysize=7.cm \epsfclipoff \fboxsep=0pt
\setlength{\unitlength}{0.7cm}
\begin{picture}(9,9)(0,0)
\put(-2,-0.5){{\epsffile{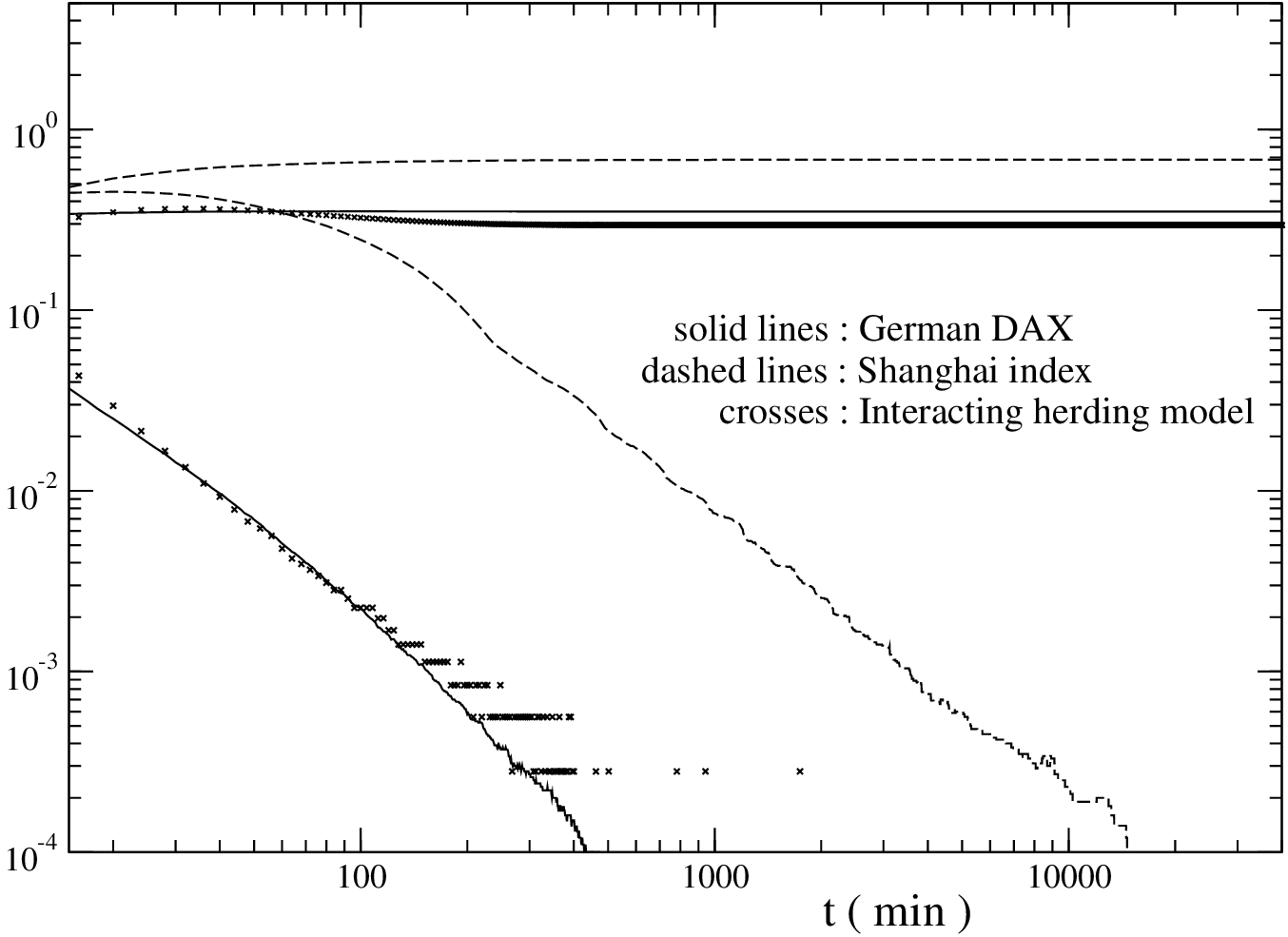}}}
\put(7,8.2){\makebox(0,0){$\partial_{z_0} P_{+}(t,z_0)|_{z_0=0}$}}
\put(4.8,3){\makebox(0,0){$\partial_{z_0} P_{-}(t,z_0)|_{z_0=0}$}}
\end{picture}
\caption{Derivatives of the persistence probabilities $P_{\pm}
(t,z_0)|_{z_0=0}$ for the minutely data in log-log scale.}
 \label{f7}
\end{figure}

Differentiation of Eq. (\ref {e20}) with respective to $z_0$ leads
to
\begin{equation}
\partial_{z_0} P_{-}(t,z_0)|_{z_0=0} \sim t^{-\theta_p+\alpha_-}.
\label{e30}
\end{equation}
In Fig. \ref {f7}, $\partial_{z_0} ~P_{-}(t,z_0)|_{z_0=0}$ has
been displayed for the minute-to-minute data of both the German
DAX and Shanghai Index. The curves decay roughly by a power law,
and the corresponding exponent $\alpha_-$ is negative for both
indices. This indicates that $z_0$ corresponds to an irrelevant
parameter around the fixed point $z_0=0$. In other words, it
induces only corrections to scaling. Corrections to scaling are
usually not universal, and so is the exponent $\alpha_-$.

\begin{figure}[p]\centering
\epsfysize=7.cm
\epsfclipoff
\fboxsep=0pt
\setlength{\unitlength}{0.7cm}
\begin{picture}(9,9)(0,0)
\put(-2,-0.5){{\epsffile{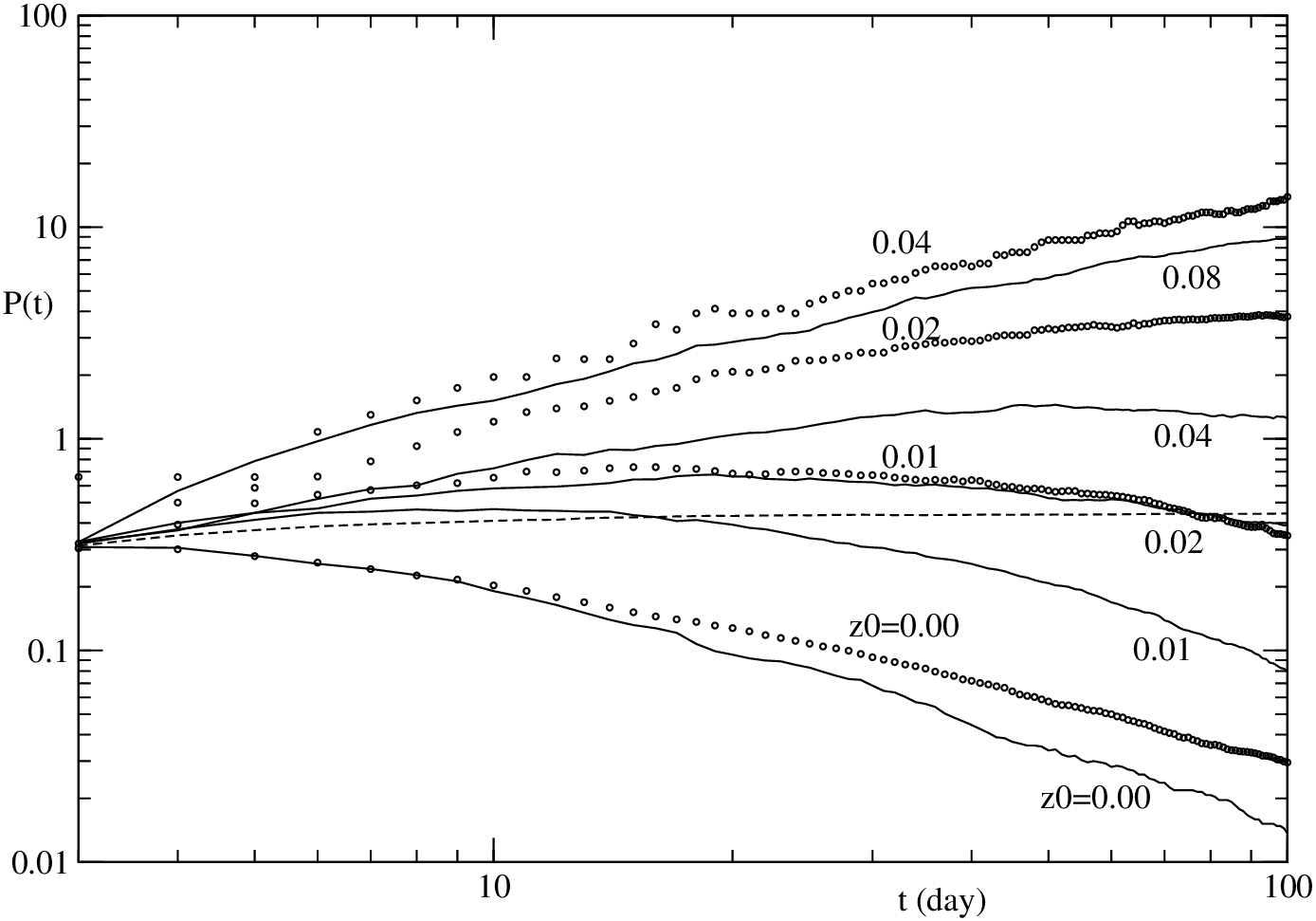}}}
\put(4.,8.5){\makebox(0,0){solid lines: $\partial_{z_0}
P_{-}(t,z_0)$ of Shanghai Index}}
\put(4.,7.7){\makebox(0,0){circle lines: $\partial_{z_0}
P_{-}(t,z_0)$ of German DAX}}
\end{picture}
\caption{Derivatives of the persistence probabilities $P_{\pm}
(t,z_0)$ for the daily data in log-log scale. The dashed line is
$\partial_{z_0} P_{+}(t,z_0)|_{z_0=0}$ of the Shanghai Index. }
 \label{f8}
\end{figure}

In Fig.~\ref {f8}, $\partial_{z_0} P_{-}(t,z_0)$ has been plotted
for different values of $z_0$ for the daily data of both the
German DAX and Shanghai Index. For big values of $z_0$, the curves
of both indices look approaching a power-law behavior with a
positive slope about $0.3$. This indicates that $z_0$ corresponds
to a relevant parameter. As $z_0$ changes from $0$ to a large
value, it occurs a crossover behavior. Similar results are also
obtained by analyzing the minute-to-minute data of both indices.
At this point, besides the scales of $z_0$, we are not able to
detect any significant difference between the German DAX and
Shanghai Index within the accuracy of our data.

For $P_{+}(t,z_0)$, we do not observe any power-law behavior at
$z_0=0$ for both indices, and the standard scaling form does not
exist. To understand the effect of $z_0$, however, we may still
suggest a 'quasi-scaling' form:
\begin{equation}
P_{+}(t,z_0) = f(t) F_{+}(g(t)z_0). \label{e60}
\end{equation}
Its derivative leads to
\begin{equation}
\partial_{z_0} P_{+}(t,z_0)|_{z_0=0} \sim f(t) g(t).
\label{e70}
\end{equation}
In Fig.~\ref {f7} and ~\ref {f8}, $\partial_{z_0}
~P_{+}(t,z_0)|_{z_0=0}$ has been displayed for the
minute-to-minute data and daily data of both indices. In all
cases, it tends to a constant, i.e., $g(t) \sim 1/f(t)$, and
$g(t)$ is an increasing function of $t$. In this sense, $z_0$
corresponds to a relevant parameter.

For comparison, we have also performed a similar analysis for the
interacting herding model. In Fig.~\ref {f7}, $\partial_{z_0}
~P_{-}(t,z_0)|_{z_0=0}$ and $\partial_{z_0}~P_{+}(t,z_0)|_{z_0=0}$
have been displayed by the cross lines. The model qualitatively
reproduces the minute-to-minute behavior of the real markets.

In the case of $z_0 \neq 0$, a similar cross-over behavior as in
Fig. \ref {f8} is also observed in the numerical simulations of
the dynamic herding model. But the slope of the curve with a big
$z_0$ is rather close to $1.0$, different from $0.3$ of the real
market.

\section{Conclusions}

In conclusions, we have investigated the persistence probabilities
$P_{\pm} (t,z_0)$ defined with the magnitude of the logarithm
price change of the financial index, using the data of the German
DAX and Shanghai Index. A power-law behavior is observed for
$P_{-} (t,z_0=0)$ up to some months for both indices. The
minute-to-minute data and daily data of the German DAX
consistently yield a persistence exponent, $\theta_p=0.89(2)$,
while the minutely data and daily data of the Shanghai Index do
not give a same persistence exponent. However, if we 'renormalize'
the minutely data of the Shanghai Index to the time scale of a
working day, i.e., calculate $Z(t',\Delta t)$ with $\Delta t = 300
\ mins$, a 'universal' persistence exponent $\theta_p=0.80(2)$ is
obtained. These results indicate that both the German DAX and
Shanghai Index are indeed long-range correlated in time, but they
very probably belong to different universality classes.

$P_{+} (t,z_0=0)$ decays much faster than $P_{-} (t,z_0=0)$ and
does not show a universal scaling behavior. This reflects a
high-low asymmetry. In addition, we find that the parameter $z_0$
around $z_0=0$ corresponds to an irrelevant parameter for $P_{-}
(t,z_0)$, while a relevant parameter for $P_{+} (t,z_0)$. The
parameter $z_0$ tends to be relevant at the big values.

We have performed numerical simulations of an interacting herding
model and a Minority Game with an inactive strategy as well as
some other models. In the case of $z_0$ around zero, the
interacting herding model nicely reproduces the scaling behavior
of the minute-to-minute data of the real markets, at least for the
German DAX. It remains challenging to simulate the dynamics of the
daily data and the behavior far away from the fixed point $z_0=0$.
Further understanding of the Shanghai Index is also needed.

Acknowledgements: Work supported in part by DFG (Germany) under
Grant No. TR 300/3-3, NNSF (China) under Grant Nos. 10325520 and
70371069.

%\bibliographystyle{/home/zheng/tex/stybase/prsty}
%\bibliography{/home/zheng/tex/stybase/eco,/home/zheng/tex/stybase/ising,/home/zheng/tex/stybase/zheng}

\end{document}